\documentclass[showkeys,floatfix]{revtex4-1}
\usepackage{graphicx}
\usepackage{amssymb}
\usepackage{amsmath}

\begin{document}
\title{Impact of mesoscale eddies on water transport between the Pacific Ocean and the Bering Sea}

\author{S.V. Prants}
\affiliation{Pacific Oceanological Institute of the Russian Academy of Sciences,\\
43 Baltiiskaya st., 690041 Vladivostok, Russia}
\homepage{htpp://dynalab.poi.dvo.ru}
\email{prants@poi.dvo.ru}
\author{A.A. Andreev}
\affiliation{Pacific Oceanological Institute of the Russian Academy of Sciences,\\
43 Baltiiskaya st., 690041 Vladivostok, Russia}
\author{M.V. Budyansky}
\affiliation{Pacific Oceanological Institute of the Russian Academy of Sciences,\\
43 Baltiiskaya st., 690041 Vladivostok, Russia}
\author{M.Yu. Uleysky}
\affiliation{Pacific Oceanological Institute of the Russian Academy of Sciences,\\
43 Baltiiskaya st., 690041 Vladivostok, Russia}

\begin{abstract}
Sea surface height anomalies observed by satellites in 1993--2012 are combined
with simulation and observations by surface drifters and Argo floats to study
water flow pattern in the Near Strait (NS) connected the Pacific Ocean with
the Bering Sea. Daily Lagrangian latitudinal maps, computed with the AVISO surface
velocity field, and calculation of the transport across the strait show that the
flow through the NS is highly variable and controlled by mesoscale and submesoscale
eddies in the area. On the seasonal scale, the flux through the western part
of the NR is negatively correlated with the flux through its eastern part ($r=-0.93$).
On the interannual time scale, a significant positive correlation
($r=0.72$) is diagnosed between the NS transport and the wind stress in winter.
Increased southward component of the wind stress decreases the northward water
transport through the strait. Positive wind stress curl over the strait area in
winter--spring generates the cyclonic circulation and thereby enhances the
southward flow in the western part ($r=-0.68$) and northward flow in the eastern
part ($r=0.61$) of the NR. In fall, the water transport in different parts
of the NS is determined by the strength of the anticyclonic mesoscale eddy
located in the Alaskan Stream area.
\end{abstract}


\maketitle

\section{Introduction}
\begin{figure}[!ht]
\begin{center}
\includegraphics[width=0.6\textwidth,clip]{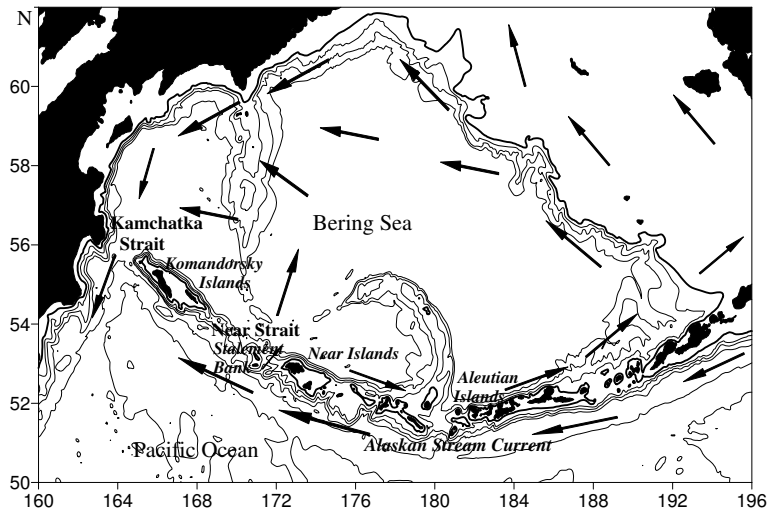}
\end{center}
\caption{Schematic representation of the currents in
the northern North Pacific and the Bering Sea.}
\label{fig1}
\end{figure}

Circulation in the Bering Sea is strongly influenced by the Alaskan Stream,
which enters the Bering Sea through the many passes in the Aleutian Arc.
The Aleutian Arc forms a porous boundary between the Bering Sea and the North Pacific.
The inflow into the Bering Sea is balanced by outflow through the Kamchatka Strait,
so that circulation in the Bering Sea basin may be described as a continuation of
the North Pacific subarctic gyre (Fig.~\ref{fig1}). Most of the transport into the
Bering Sea occurs through the Near Strait ($\text{6--12} \times 10^6$~m${}^3$/s). The NS with its maximum
depth of $1600$~m, averaged depth of $650$~m and width of $360$~km is located between the Komandorsky
Islands and the Attu Island (Near Islands). The existence of the Stalemate Bank
(near $53^{\circ}$N, $171^{\circ}$E), with depths $<100$~m, in the eastern deep part of the NS may
lead to topographically-induced dynamics in strait flows. Recent studies (current
mooring measurements and model simulations) indicate that the flow through the NS, as
through the other Aleutian passes, is extremely variable
\citep{Reed_Stabeno_1993, Ezer_Oey_2013, Clement_Kinney_Maslowski_2012, uk_Luchin_Nechaev_Kukuchi_2012}.
Occasionally, instabilities (eddies) occur in the Alaskan Stream that inhibit flow into the Bering Sea
through the NS \citep{Stabeno_Reed_1992}. The data from current moorings demonstrated that the inflow
through the NS is not to be bounded and may migrate laterally \citep{Stabeno_Reed_1992}.
Ezer and Oey \citep{Ezer_Oey_2013} stated that the NS (``wide passage''), does not have well
organized currents, but they are largely influenced by the mesoscale eddies that propagated across the strait.
Based on modeling results, Kinney and Maslowski \citep{Clement_Kinney_Maslowski_2012} concluded that the
interannual variability in mass transport and property fluxes is particularly strong across the
western Aleutian Island Passes, including the NS. Much of this variability can be
attributed to the presence of meanders and eddies found both to the north and south
of the passes, which are found to directly cause periodic flow reversals and maxima
in the western passes. Panteleev \citep{uk_Luchin_Nechaev_Kukuchi_2012} combined the sea surface
height anomalies (observed by satellites in 1992--2010) with monthly climatology of
temperature and salinity to estimate circulation in the southern Bering Sea.
On the interannual time scale, significant negative correlation ($r=-0.84$)
was diagnosed between the NS transport and the Alaskan Stream. In particular,
the NS inflow amplifies when the upstream Alaskan Stream is relatively weak,
resulting in stronger recirculation in the western Bering Sea. Stronger
Alaskan Stream appears to reduce the NS inflow and at the same time produces
larger transport through the Aleutian Arc, which amplifies the main cyclonic
gyre controlled by the continental slope within the Bering Slope Current
region \citep{uk_Luchin_Nechaev_Kukuchi_2012}.

The Alaskan Stream carries relatively warm subsurface waters. Eddies in the Alaskan Stream
along the Aleutian Islands could have a significant impact on the heat, freshwater, nutrient,
and biota exchange between the coastal area to the south of the Aleutian Islands and the offshore
region in the western and central subarctic North Pacific. In addition, eddies to the south of the
Aleutian Islands were suggested to drive the flow between the North Pacific and the Bering Sea
\citep{Okkonen_1996}. The  effect  of continued  disruption  of inflow  to  the  Bering Sea
would be an  appreciable  cooling  ($0.5$ ${}^{\circ}$C) of subsurface waters
\citep{Stabeno99}. These substantial changes in seawater temperature would
affect the climate and ecosystem of the Bering Sea.

The goal of the paper is two-fold.
Firstly, we use the NS as a case study to understand general characteristics of the strait
dynamics. The second aim is to shed light on impact of mesoscale eddies on the water flux
between the North Pacific and the Bering Sea.

We work in the framework of the Lagrangian
approach to study transport at the sea surface in which one follows fluid particle trajectories
in a velocity field calculated from altimetric measurements or obtained as an output of one
of the ocean circulation models. The Lagrangian approach has been successfully applied to
study large-scale mixing and transport in different seas and oceans (see, for example,
Refs.~\citep{H02,Abraham02,OF04,KP06,Olascoaga06,LO07,Beron08,Harrison10,OM11,Keating11,
Hernandez11,Huhn12,P13,FAO13} and references therein).
In these papers methods, borrowed from dynamical systems theory, have been used to
find hidden geometric structures in the ocean that organize complicated large-scale
flows. Ones of the most important among them are Lagrangian coherent structures which
are proxies of so-called stable and unstable manifolds of hyperbolic  trajectories of
fluid particles. Hyperbolicity is a concept in dynamical systems theory (for a review
on application of that theory in physical oceanography see \citep{Wiggins,KP06}).
It is characterized by the presence of expanding and contracting directions in the
phase space of a dynamical system. This is a situation where phase trajectories converge
in one direction and diverge in the other one. A hyperbolic stagnation point in a stationary
system has stable and unstable manifolds along which the phase points approach it and move
away from it, respectively. Typically, under a perturbation, a hyperbolic stagnation point
of the unperturbed system becomes a hyperbolic trajectory, $\gamma(t)$, with stable,
$W_s(\gamma)$, and unstable $W_u(\gamma)$ invariant manifolds. In the enlarged phase
space those manifolds are surfaces which do not coincide but intersect each other.
A blob of the phase fluid in the phase space experiences stretching and folding that
typically gives rise to chaotic motions even in purely deterministic dynamical systems,
the remarkable phenomenon known as deterministic or dynamical chaos. The ultimate reason
of that chaos is the instability of trajectories that is expressed in terms of the
hyperbolicity conditions.

In fluid flows invariant manifolds are material surfaces composing of the same fluid
particles in the course of time. In 2D fluid flows stable  and unstable manifolds of
a hyperbolic trajectory $\gamma(t)$~are material lines consisting of a set of points
through which at time moment $t$ pass trajectories asymptotically to $\gamma(t)$ at
$t \to \infty$ ($W_s$) and $t \to -\infty$ ($W_u$) \citep{Wiggins,KP06}. They are
complicated curves infinite in time and space that act as boundaries to fluid transport.
Theory of chaotic motions, known as chaotic advection \citep{Aref}, is well developed in
time periodic 2D flows where periodic hyperbolic trajectories can be seen as fixed points
using Poincar{\'e} maps. The stable and unstable manifolds of hyperbolic trajectories can
be computed using different techniques and are known to form complex homoclinic or
heteroclinic tangles which are ``seeds'' of chaotic advection.

There is a commonly used method for detecting stable and unstable manifolds in irregular and
even turbulent flows based on computing Lyapunov exponents where a region under study is
seeded with a large number of tracers for each of which the finite-time Lyapunov exponents
are computed forward and backward in time. The curves of local maxima (ridges) in the
Lyapunov scalar field, attributed to initial tracer's positions, approximate stable manifolds
when solving equations of motion forward in time and unstable ones when solving them backward
in time \citep{Abraham02,Beron08,OF04,LO07,OM11,FAO13}.

We apply the Lagrangian approach to model transport through the NS and compute trajectories of
a large number of synthetic tracers advected by the interpolated AVISO geostrophic velocity field.
Our main goal is to study not mixing but transport through the NS that can be characterized more
adequately by computing not the Lyapunov exponents but the other Lagrangian indicators that have
been recently introduced in \citep{OM11,DAN11,P13,FAO13}. In the context of this paper, the most appropriate among them are meridional displacements of
tracers that are computed after solving the advection equations backward in time in an interpolated
altimetric velocity field. As the result, we get daily synoptic latitudinal maps around
the NS that show clearly not only mesoscale and submesoscale eddies in the region but as well origin and
history of waters crossing the strait for a given period of time.

The paper is organized as follows. Section~2 details the data and methods used in the paper.
Section~3 contains the main results. We consider typical hydrological situations in different seasons
and years with various types of eddies to be present in the strait and around it. The representative
latitudinal  synoptic maps demonstrate clearly those eddies and  provide us with information about the
direction of transport through different parts of the strait and its  dependence on the number and
polarity of the eddies. The daily transport across different parts of
the strait has been calculated in 1993--2012 to show that the flow is highly variable and controlled
by mesoscale and submesoscale eddies in the area. The section is ended with establishing a correlation
between the NS transport and the wind stress.

\section{Data and methods}

Since 1992, the ocean surface topography has been continuously observed from space by the Topex/Poseidon,
European Remote Sensing (ERS), Geosat Follow-On (GFO), Envisat and Jason satellite missions. These data
are available at http://www.aviso.oceanobs.com. We use the gridded sea surface height anomalies (SSHA)
for the period of 1993--2012 for the diagnostic computations. For each $7$-day period, the SSHAs,
obtained by optimal interpolation of all the available altimeter missions, were downloaded from
the Aviso site.

Geostrophic velocities  were obtained from the AVISO database. The data is gridded on a
$1/3^{\circ}\times 1/3^{\circ}$ Mercator grid. Geostrophic velocities were calculated from the
sea-surface height anomalies by solving the equation for geostrophic equilibrium. Bicubical
spatial interpolation and third order Lagrangian polynomials in time have been used to provide
accurate numerical results. Lagrangian trajectories have been computed by integrating the advection
equations with a fourth-order Runge-Kutta scheme with a fixed time step of $0.001$th part of a day.

Motion of a fluid particle in a two-dimensional flow is the trajectory of a dynamical system with
given initial conditions governed by the velocity field. The corresponding advection equations
are written as follows:
\begin{equation}
\frac{d x}{d t}= u(x,y,t),\qquad \frac{d y}{d t}= v(x,y,t),
\label{adveq}
\end{equation}
where the longitude, $x$, and the latitude, $y$, of a passive particle are in geographical minutes,
$u$ and $v$ are angular zonal and meridional components of the surface velocity expressed in minutes per day.

Drifter data were used in the study for validation purposes and consisted of two major components:
surface drifters and subsurface Argo buoys. In this study, we use the monthly wind stress dataset
from the NCEP reanalysis \citep{Saha_White_Woollen_et_al__1996}. The horizontal resolution of the NCEP data is
$1.9^{\circ}\,\text{lat}\times 1.8^{\circ}\,\text{lon}$.

Some our conclusions are based on locating eddies on the Lagrangian synoptic latitudinal  maps computed
with the help of the altimetric geostrophic velocity field. The boundaries of the eddies are, as a
rule, surrounded by the stable and unstable invariant manifolds
of hyperbolic trajectories located nearby. Those structures to be computed at the ocean surface are
determined mainly by the large-scale advection field  which is appropriately captured by altimetry.
The question is how robust are those Lagrangian features to inevitable errors in altimetric velocity
fields. That problem has been carefully studied in literature. It has been shown theoretically
\citep{H02} that the strong stable and unstable manifolds, existing for a sufficient long time,
are robust to errors in observational or model velocity fields because the particle trajectories
will in general diverge exponentially from the true trajectories near a stable manifold.
The very manifolds are not expected to be perturbed to the same degree, because errors in the
particle trajectories spread along them. They were found to be relatively insensitive to both
sparse spatial and temporal resolution and to the velocity field interpolation method
\citep{Harrison10, Keating11, Hernandez11}. In particular, the method for locating boundaries
of large eddies (more than $20$~km in diameter) and strong jets with the help of Lyapunov
exponents has been shown to be reliable \citep{Abraham02,Beron08,OF04,LO07,OM11,P13,FAO13}.
However, small Lagrangian features are not well resolved from altimetry and should be
considered with some caution. The comparison of the  Lagrangian coherent structures computed
with altimetric velocity fields in numerous paper (see, for example,
\citep{Abraham02,Olascoaga06,Beron08,Ovidio09,Huhn12}) with independent satellite and
in situ measurements of thermal and other fronts in different seas and oceans
has been shown a good correspondence.

\section{Results and Discussion}
\subsection{Calculation of the water transport across the Near Strait}

The width of the NS is sufficient (greater than the internal Rossby radius $\sim 10$--$20$~km)
to provide a northward flow on the east side and a southward flow on the west side.
The relative strength of these two branches determines the direction and magnitude of
the net transport in the NS that has been computed as follows. The line, crossing the
strait from $168.5^{\circ}$E,~$54.5^{\circ}$N to $172.3^{\circ}$E,~$53^{\circ}$N,
consists of 20 segments with 100 particles to
be placed at each segment. The daily transport across each segment may be obtained by
integrating velocity normal at a given point across the segment length.
The northward (southward) transport is supposed to be positive (negative).

\begin{figure}[!ht]
\begin{center}
\includegraphics[width=0.5\textwidth,clip]{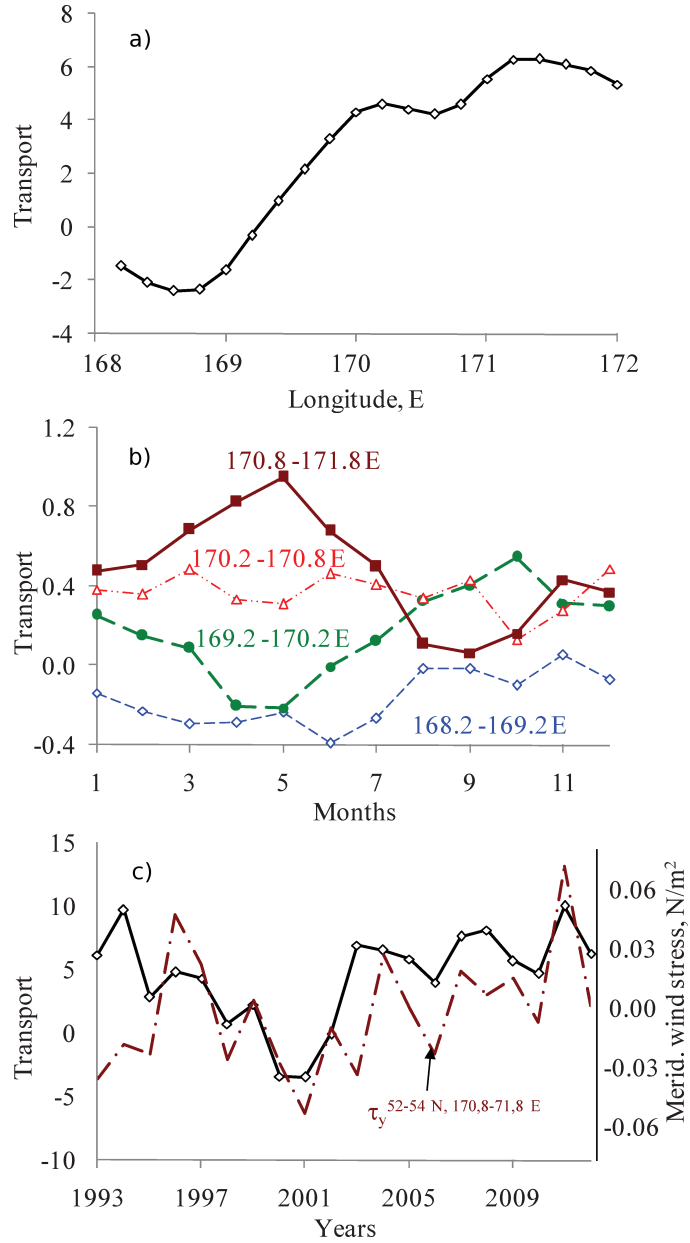}
\end{center}
\caption{Annually averaged water transport in the NS versus
longitude (a); monthly averaged water transport in different
parts of the NS (b) and the year-to-year changes of the water
transport in the NS and the meridional wind stress (Nov.--Mar.)
in the NS area (c).}
\label{fig2}
\end{figure}

Fig.~\ref{fig2}a shows that during the study period (1993--2012~yrs) the surface northward
flow was enhanced in the central and eastern parts of the strait but the southward flow in
the western side of strait was weak. The net transport through the NS was directed from the
North Pacific to the Bering Sea. The transport time series through the western and eastern
NS are negatively correlated (at the $99\%$ significance level) (Fig.~\ref{fig2}b).
The correlation coefficient is $-0.93$ for the monthly mean values. The correlation essentially means that when
the inflow into the Bering Sea via the eastern NS is strong, the outflow via the western NS
tends to be weak or southward. There is seasonality in the water flux through the NS.
The northward water transport through the central and eastern parts of the NS is strong between
March and June and relatively weak in August--October. The year-to-year change of the
annually averaged volume transport through the NS is significantly correlated with the
meridional wind stress ($\tau_y$, November--March) (Fig.~\ref{fig2}c).
An increase of southward component of the wind stress (positive values of $\tau_y$)
decreases the northward water flux through the strait ($r=-0.72$, 1997--2012).

\subsection{Lagrangian synoptic maps}

\begin{figure}[!ht]
\begin{center}
\includegraphics[width=0.36\textwidth,clip]{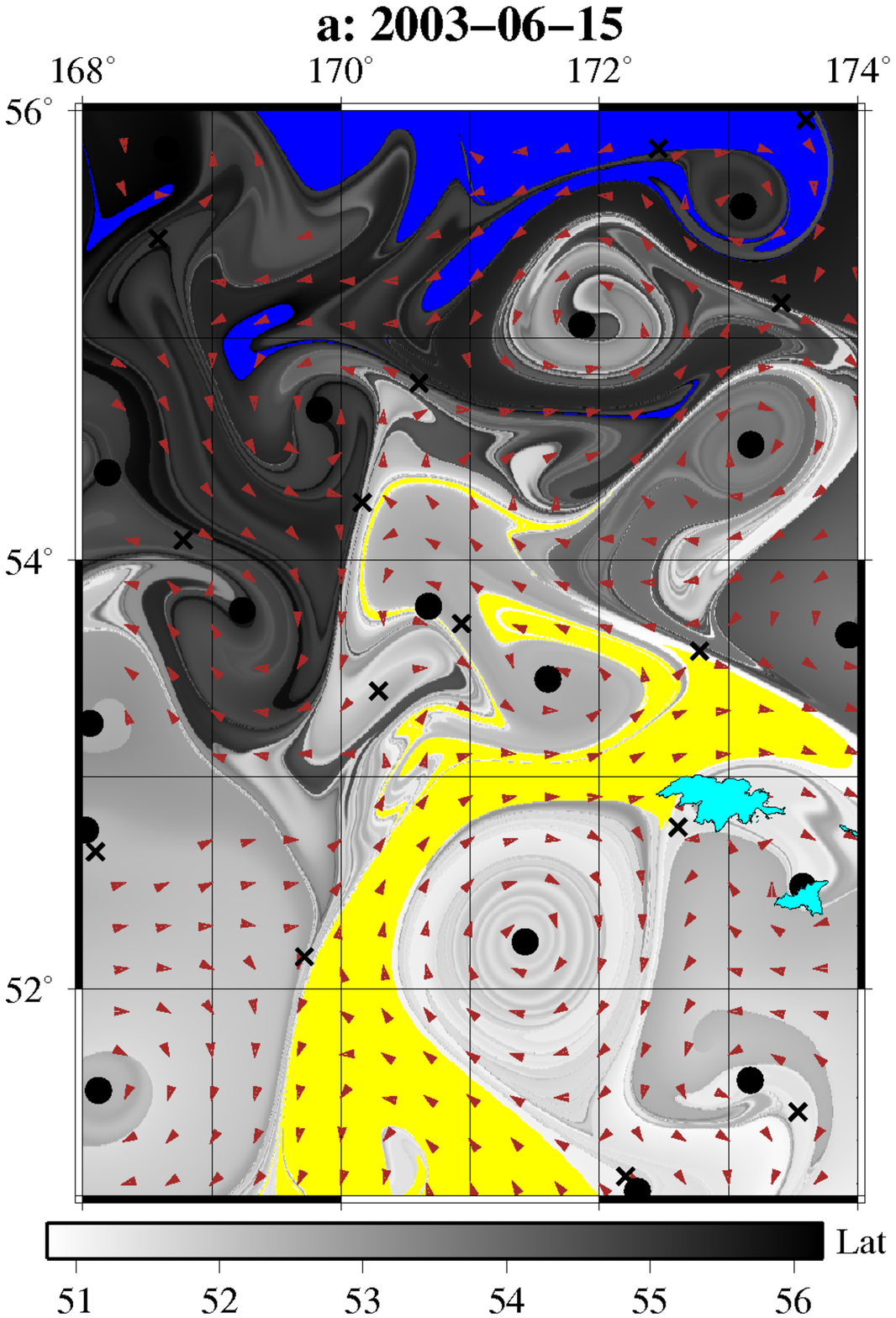}
\includegraphics[width=0.36\textwidth,clip]{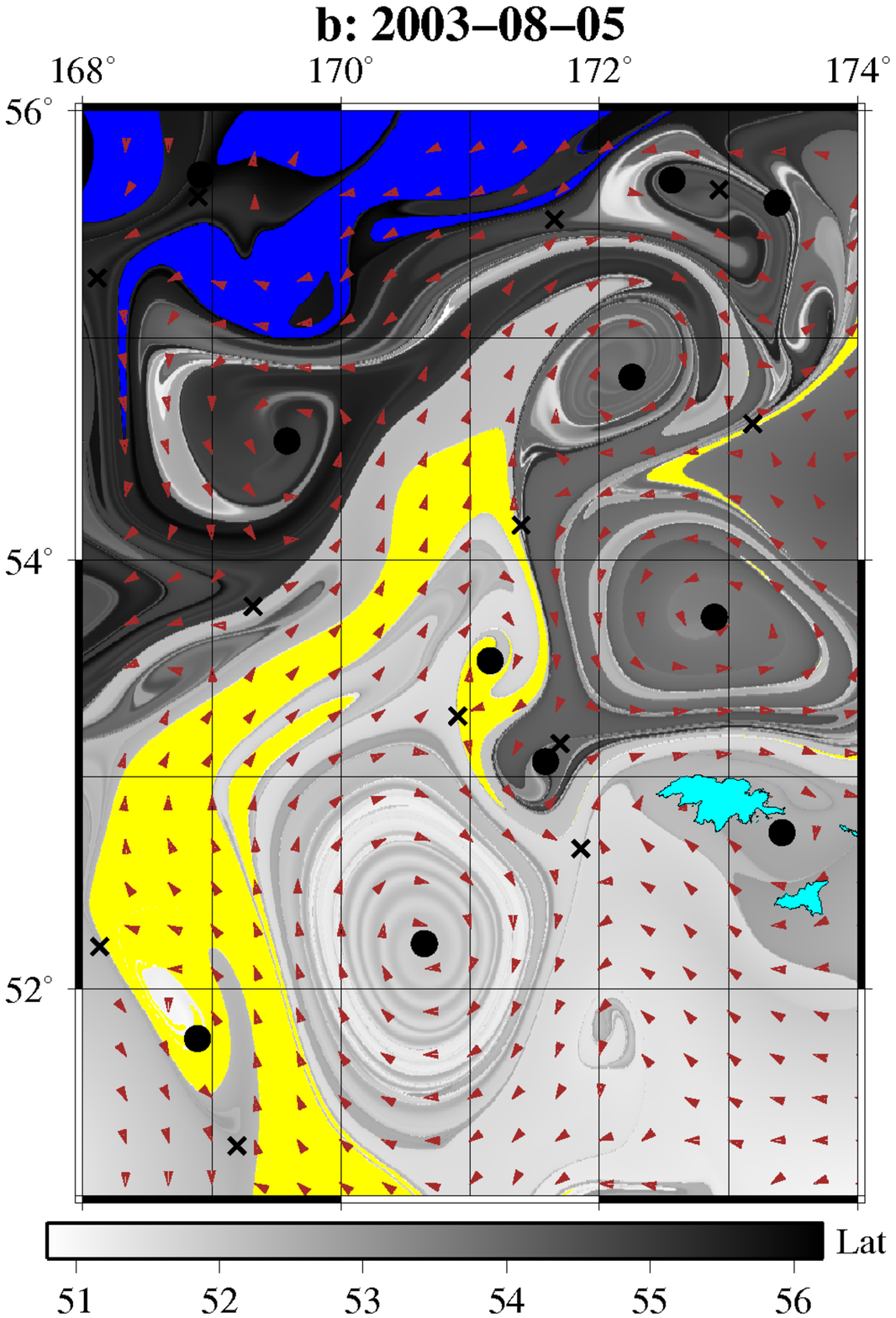}\\
\includegraphics[width=0.36\textwidth,clip]{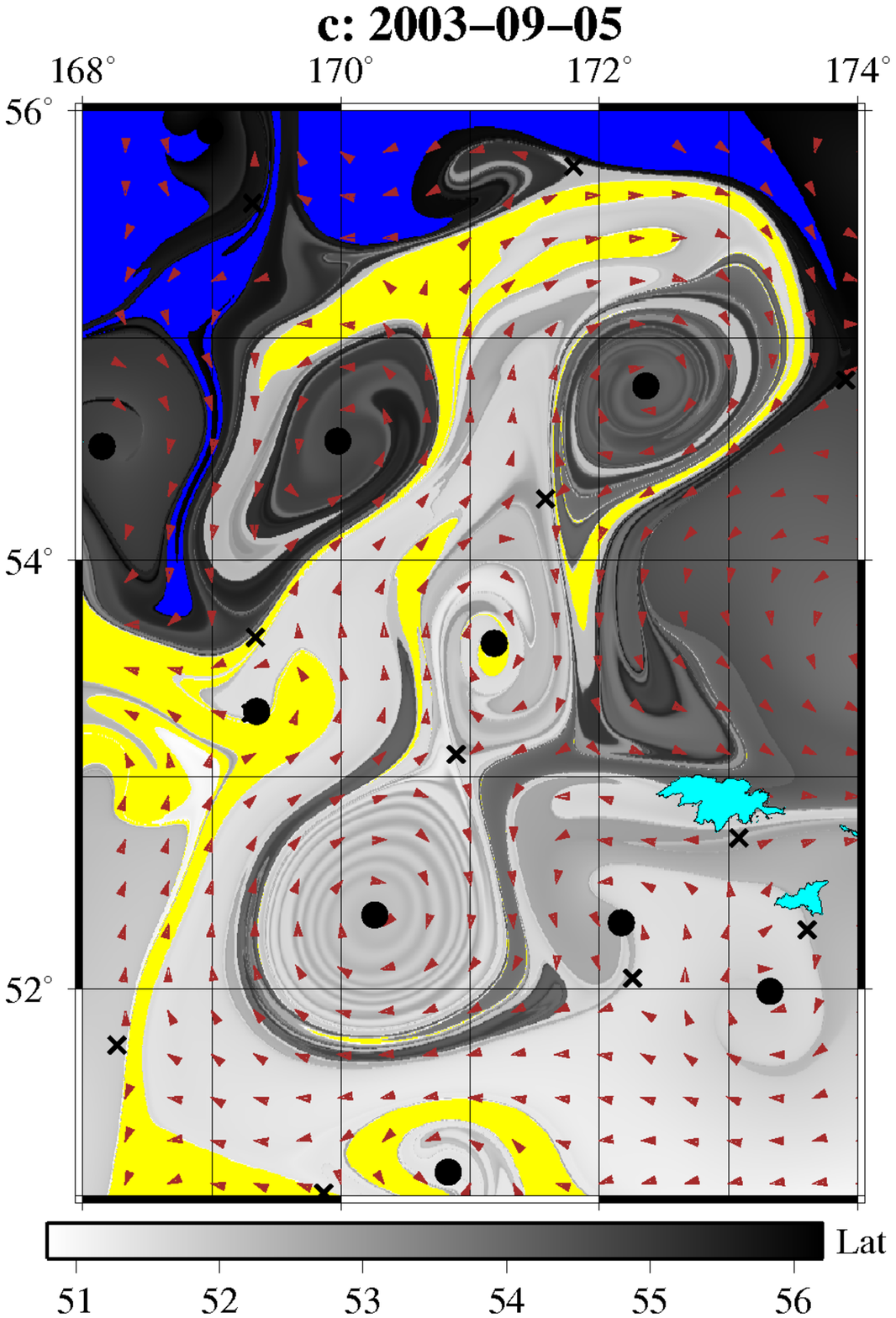}
\includegraphics[width=0.36\textwidth,clip]{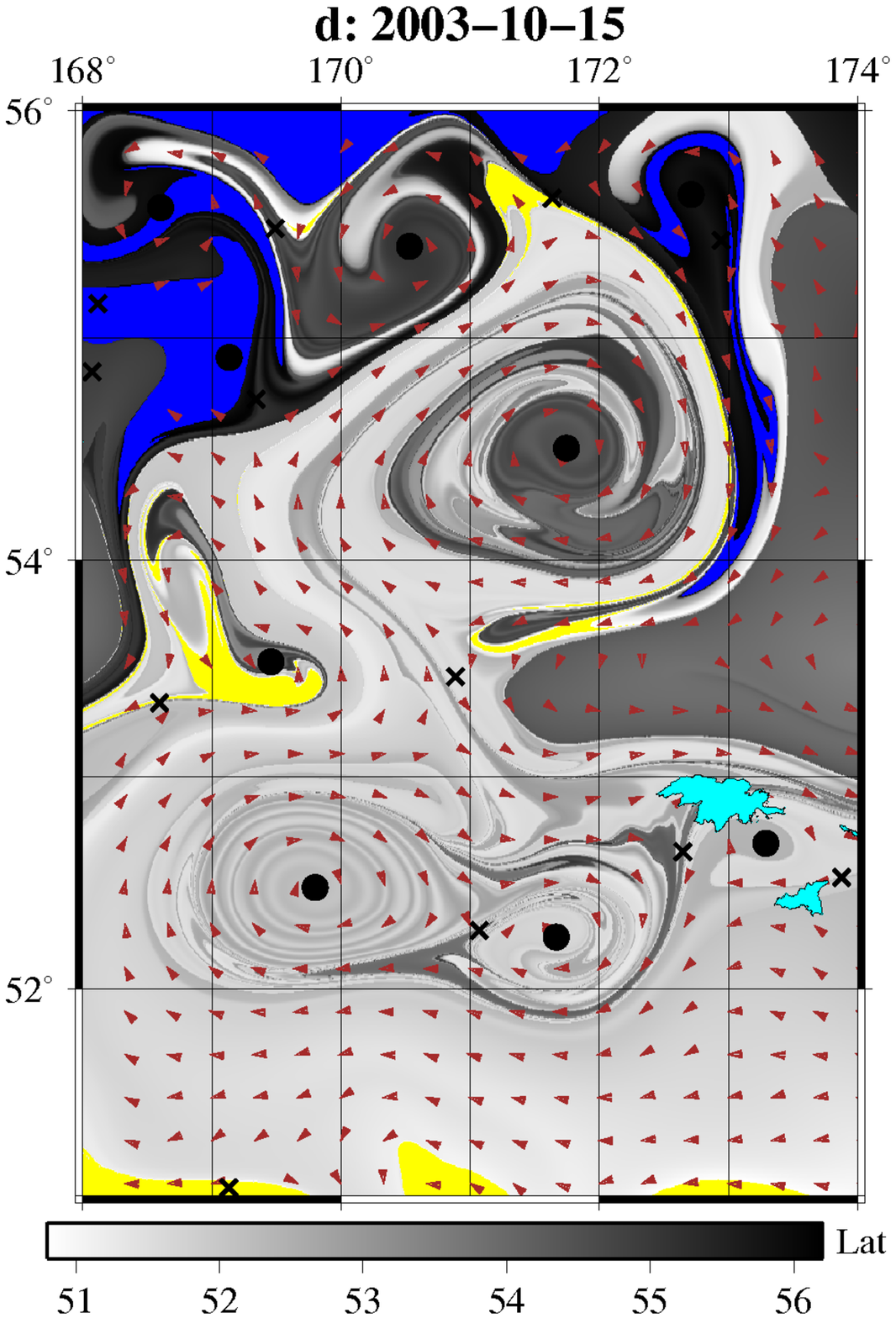}
\end{center}
\caption{Lagrangian latitudinal maps of the region in Summer--Fall of
2003 demonstrate the impact of  anticyclonic eddies on the water transport
across the NS.
The anticyclonic mesoscale eddy is located nearby the eastern part of the
strait to the
south from the strait (a); the submesoscale topographic anticyclonic eddy
is formed over
the Stalement Bank in the strait (b); triple anticyclonic vortex structure
with the
mesoscale eddy to the south of the strait, the topographic eddy over the
bank and
the mesoscale eddy to the north of the strait controls transport across
the eastern
part of the strait (c); the triple  vortex structure disappears opening
a way to the northward transport exclusively (d).
Dark grey and black colors mean that the corresponding particles came
to their positions for 90 days from the latitudes to the north of the strait
($>54^{\circ}$N), whereas the light
grey and white colors mean that the corresponding particles came from the
latitudes to the south of the
strait ($<54^{\circ}$N).
(In the web version of this article ``yellow''and ``blue'' waters crossed, respectively,
the southern and the northern boundaries of the box when moving for 90 days.)}
\label{fig3}
\end{figure}

The satellite altimetry records and buoy and drifter tracks demonstrate that anticyclonic
and cyclonic eddies to the north and south of the NS have been regularly observed.
Synoptic latitudinal maps (Fig.~\ref{fig3}) demonstrate the eddies and the origin of waters
in study area during Summer--Fall 2003.
All the maps have been computed by integrating the advection equations (\ref{adveq}) for a
large number of particles backward in time for 90 days before the date indicated on the
corresponding map and coding by color the latitudes from which the corresponding particles
came to their final positions on the map. Such maps provide information on the advection
of waters through the strait in the meridional direction.

We mark ``instantaneous'' hyperbolic and elliptic stagnation points on the overlaid AVISO
velocity field  by crosses and circles, respectively. The elliptic points are usually situated
in the centers of eddies and gyres. It is well known that besides ``trivial'' elliptic points,
the motion around which is stable, there are hyperbolic  points which organize fluid motion in
their neighbourhood in a specific way. In a steady flow the hyperbolic points are typically
connected by the separatrices which are their stable and unstable invariant manifolds.
In a time-periodic flow the hyperbolic points are replaced by the corresponding hyperbolic
trajectories with associated invariant manifolds which in general intersect transversally
resulting in a complex manifold structure known as heteroclinic tangles. The fluid motion in
these regions is so complicated that it may be strictly called chaotic, the phenomenon known as
chaotic advection (for a review see \citep{Aref,Ottino,KP06}). Adjacent  fluid particles in such
tangles rapidly diverge providing very effective mechanism for water mixing.

Firstly, we focus at the anticyclonic eddy that was tracked in the Alaskan Stream area south
of the NS in the Spring-Fall 2003 (Fig.~\ref{fig3}a). The elliptic point (dark circle) is situated in its center
around $x_0=171^\circ 30'$E, $y_0=52^{\circ}15'$N.
Dark grey and black colors on the latitudinal  maps mean that the corresponding particles came to their positions
for 90 days from the latitudinal  to the north of the strait ($>53$--$54^{\circ}$N), whereas the light
grey and white colors mean that the corresponding particles came from the latitudinal  to the south of the
strait ($<53$--$54^{\circ}$N). It had an elliptical shape with a
dimension of $120\times 150$~km, and its center was located over the axis of the Aleutian bottom
trench. According to the altimetry data, the eddy was formed in March--April 2003 to the southeastward
of the Near Islands ($\sim 173^\circ$E). In June 2003 it occupied the position to the south of the NS.
The Lagrangian maps computed for May--July show that the northward
transport prevails in the central and eastern parts of the NS (see ``white'' and ``yellow'' tongues in
the printed and web versions of Fig.~\ref{fig3}a, respectively).

The interaction of the mesoscale eddy with the Stalement Bank led to formation of
the submesoscale topographic anticyclonic eddy (horizontal dimension of $30\times 40$~km) to
the north of the bank in early August 2003. The elliptic point of that eddy is situated in its center
around $x_0=171^\circ $E, $y_0=53^{\circ}30'$N (Fig.~\ref{fig3}b). By the beginning of September,
that submesoscale eddy was interacting with the mesoscale eddies to the north and south
of the NS forming the triple vortex structure with two hyperbolic points (crosses) between the eddies
and three elliptic ones in their centers (Fig.~\ref{fig3}c). This structure changed significantly
the character of transport, enhancing the northward flux through the central part of the NS
and southward flux on the eastern side of the NS in August--September 2003.
That result has been confirmed by direct calculation of daily and monthly fluxes in different
parts of the NS. The grey filaments around the periphery of the eddies in that structure mark
the waters that crossed for 90 days the eastern side of the NS from the north to the south.
In the language of dynamical systems theory, the corresponding fluid particles were moving
along the unstable manifolds of the hyperbolic points between the eddies that can be visualized
approximately by computing the backward-in-time finite time Lyapunov exponents.
By the end of September, the topographic eddy over the bank disappeared under the pitchfork
bifurcation that opened a way to a northward flux through the central and eastern parts
of the NS (Fig.~\ref{fig3}d). Instead of that, one can see in Fig.~\ref{fig3}d a pair of anticyclones
to be formed in October to the south of the NS with a hyperbolic point between them.

\begin{figure}[!htb]
\begin{center}
\includegraphics[width=0.35\textwidth,clip]{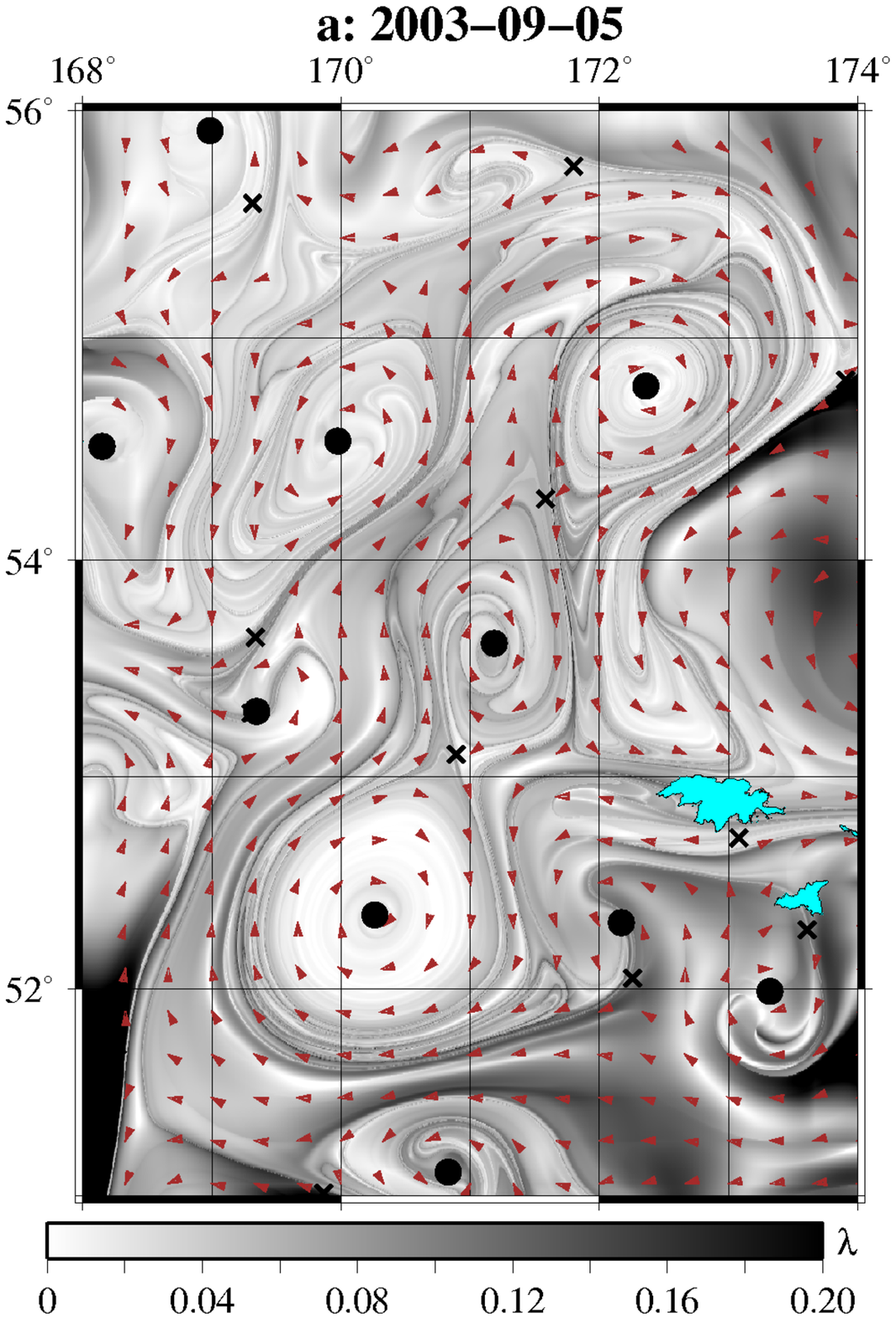}
\includegraphics[width=0.35\textwidth,clip]{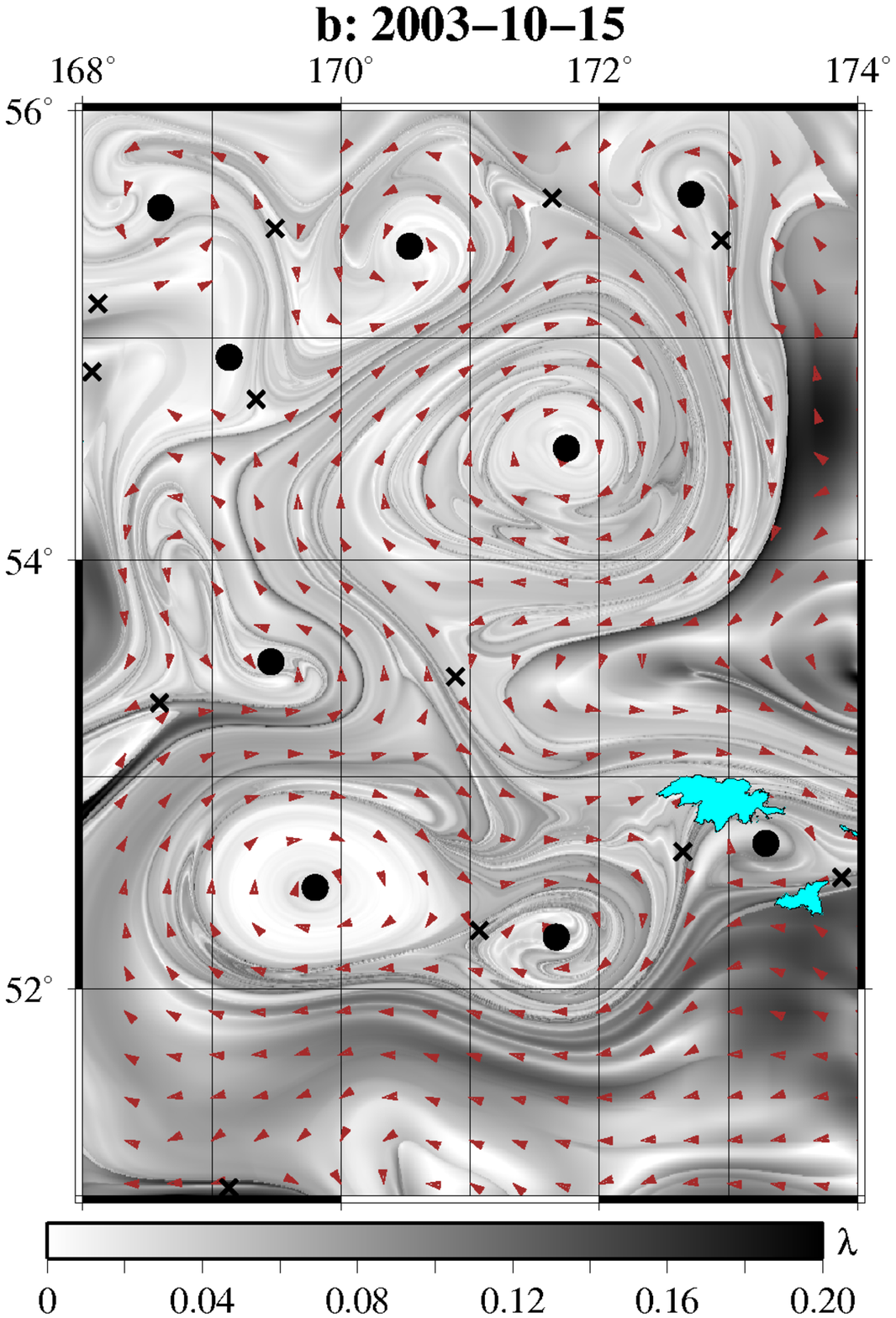}
\end{center}
\caption{Maps of the finite-time maximal Lyapunov exponents $\lambda$ (a and b) computed on the
same days as in panels c and d in Fig.~\ref{fig3} and under the same
other conditions. Color codes the values of $\lambda$ in days${}^{-1}$.}
\label{fig4}
\end{figure}

We have checked our results shown in the latitudinal  maps (Fig.~\ref{fig3}) by computing the
finite-time Lyapunov exponents using the method of the singular-value decomposition of
an evolution matrix for the linearized advection equations proposed in \citep{OM11}.
This quantity characterizes asymptotically the average rate of the particle's dispersion
and is given by formulae
\begin{equation}
\lambda(t,t_0)=\frac{\ln\sigma(t,t_0)}{t-t_0},
\label{lyap}
\end{equation}
which is the ratio of the logarithm of the maximal possible stretching in a given direction
to a time interval $t-t_0$. Here $\sigma(t,t_0)$ is the maximal singular value of the evolution matrix.
The Lyapunov field characterizes quantitatively mixing along with directions of maximal stretching
and contracting, and it is applied to identify Lagrangian coherent structures in irregular velocity fields.
In Fig.~\ref{fig4}a and b we show two Lyapunov maps on the same days as in panels c and d in Fig.~\ref{fig3},
respectively. The values of $\lambda$ have been computed by integrating the
advection equations (\ref{adveq}) backward in time for 90 days. The color in Fig.~\ref{fig4} codes the values of
$\lambda$ which are measured in days${}^{-1}$.
The black ``ridges'' (curves of the local maxima) of the Lyapunov field are known to delineate stable manifolds
of the most influential hyperbolic trajectories in the region when integrating advection equations forward in time
and unstable ones when integrating them backward in time. The black ``ridges'' on the map in Fig.~\ref{fig4}
delineate the corresponding unstable manifolds which are by definition the curves of maximal stretching.

Figure~\ref{fig4}a on September 5, 2003 should be compared with Fig.~\ref{fig3}c on the same day.
Both the figures demonstrate clearly a triple vortex structure, consisting of three anticyclonic eddies,
with elliptic points in their centers and two hyperbolic points between the eddies.
The black ``ridges'' in  Fig.~\ref{fig4}a,  encompassing the vortex structure, are proxies of the unstable
manifolds of those hyperbolic points. It is seen in Fig.~\ref{fig4}b on October~15, 2003, to be compared
with Fig.~\ref{fig3}d, that the topographic eddy in the NS over the Stalement Bank disappeared.
Both the Lyapunov and latitudinal  maps reveal clearly the same eddies in the area. The Lyapunov maps are
useful to characterise quantitatively mixing in the region and to reveal stable and unstable manifolds,
but they do not provide us with information about the direction of transport of water masses which is the
thing we are most interested in here. On the other hand, latitudinal  maps, providing such an information,
are not suitable to approximate those manifolds that may be of interest in studying water mixing.
Thus, the Lyapunov and latitudinal maps and synoptic maps of the other Lagrangian indicators may
serve complimentary tools to monitor transport and mixing in the ocean.

\begin{figure}[!htb]
\begin{center}
\includegraphics[width=0.6\textwidth,clip]{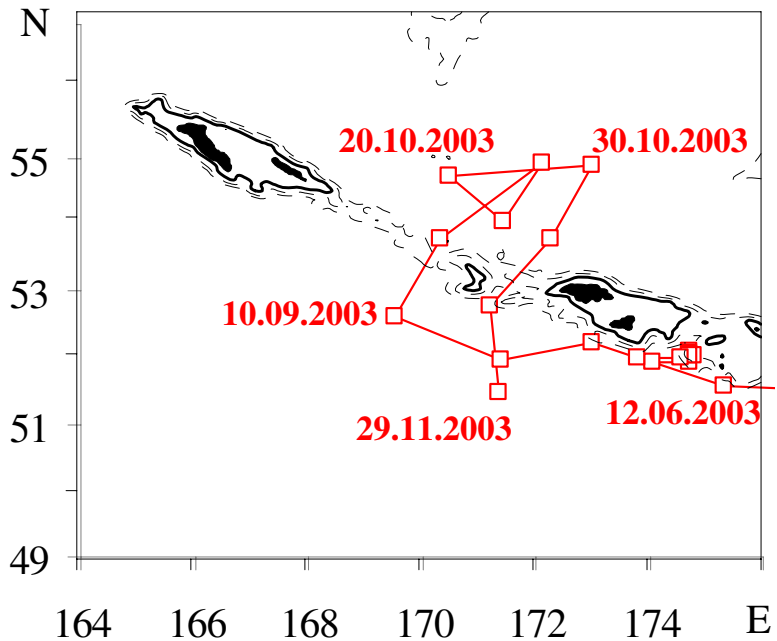}
\end{center}
\caption{Trajectory of the Argo buoy (no.~4900162) in Summer--Fall of 2003.}
\label{fig5}
\end{figure}

The simulation performed above is supported by the Argo buoy track (no.~4900162) in
Summer--Fall 2003 (Fig.~\ref{fig5}a). That buoy was captured by the anticyclonic mesoscale eddy in the
Alaskan Stream shown in Fig.~\ref{fig3}a, then exited the eddy to the south of the NS, crossed the strait
along the unstable manifold (see the grey filament encompassing that eddy on its western side in
Fig.~\ref{fig3}b). After that, it circled the anticyclonic eddy to the north of the NS (Figs.~\ref{fig3}c
and d and \ref{fig4}a and b) and then crossed the NS through the eastern part of the strait.

\begin{figure}[!htb]
\begin{center}
\includegraphics[width=0.38\textwidth,clip]{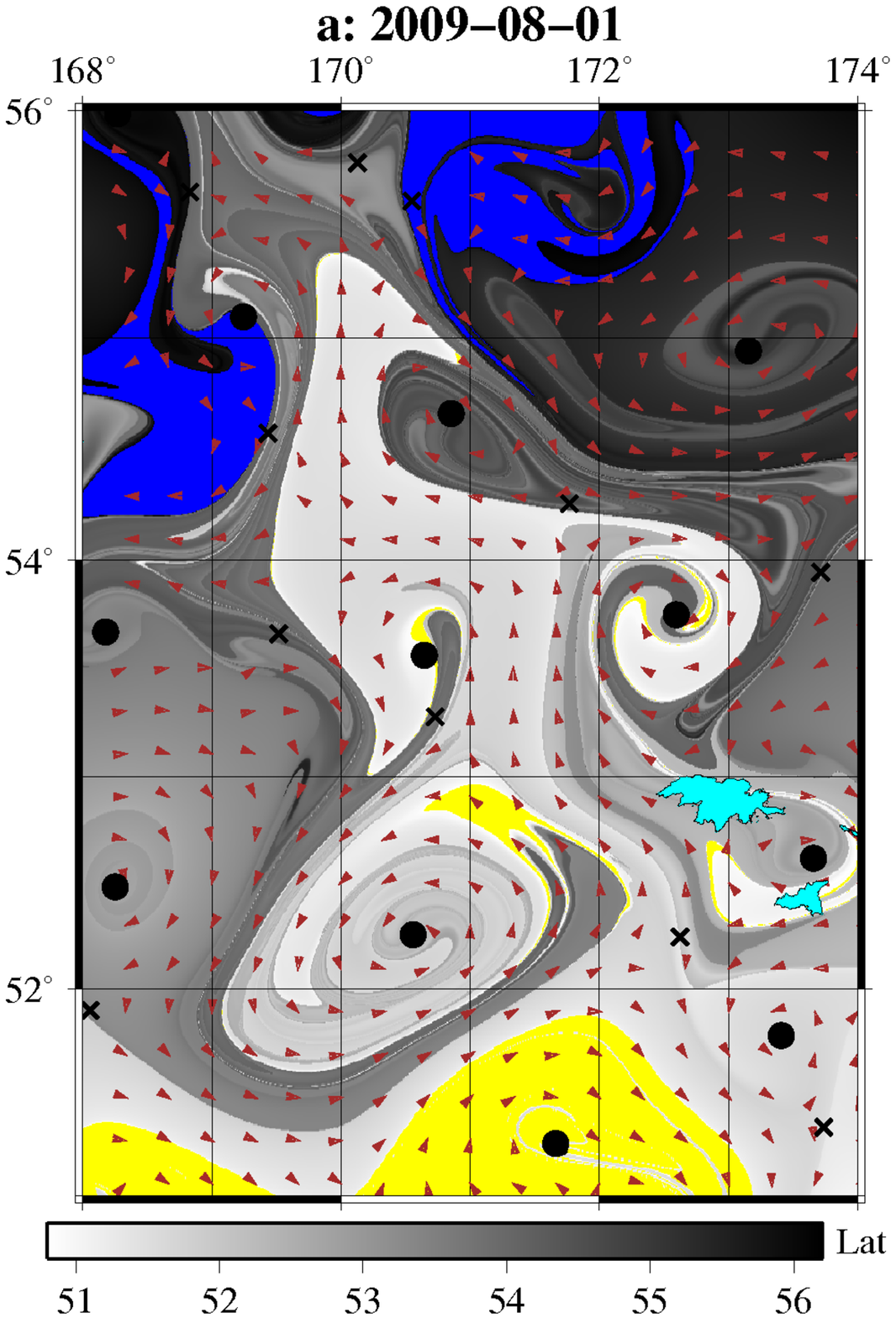}
\includegraphics[width=0.38\textwidth,clip]{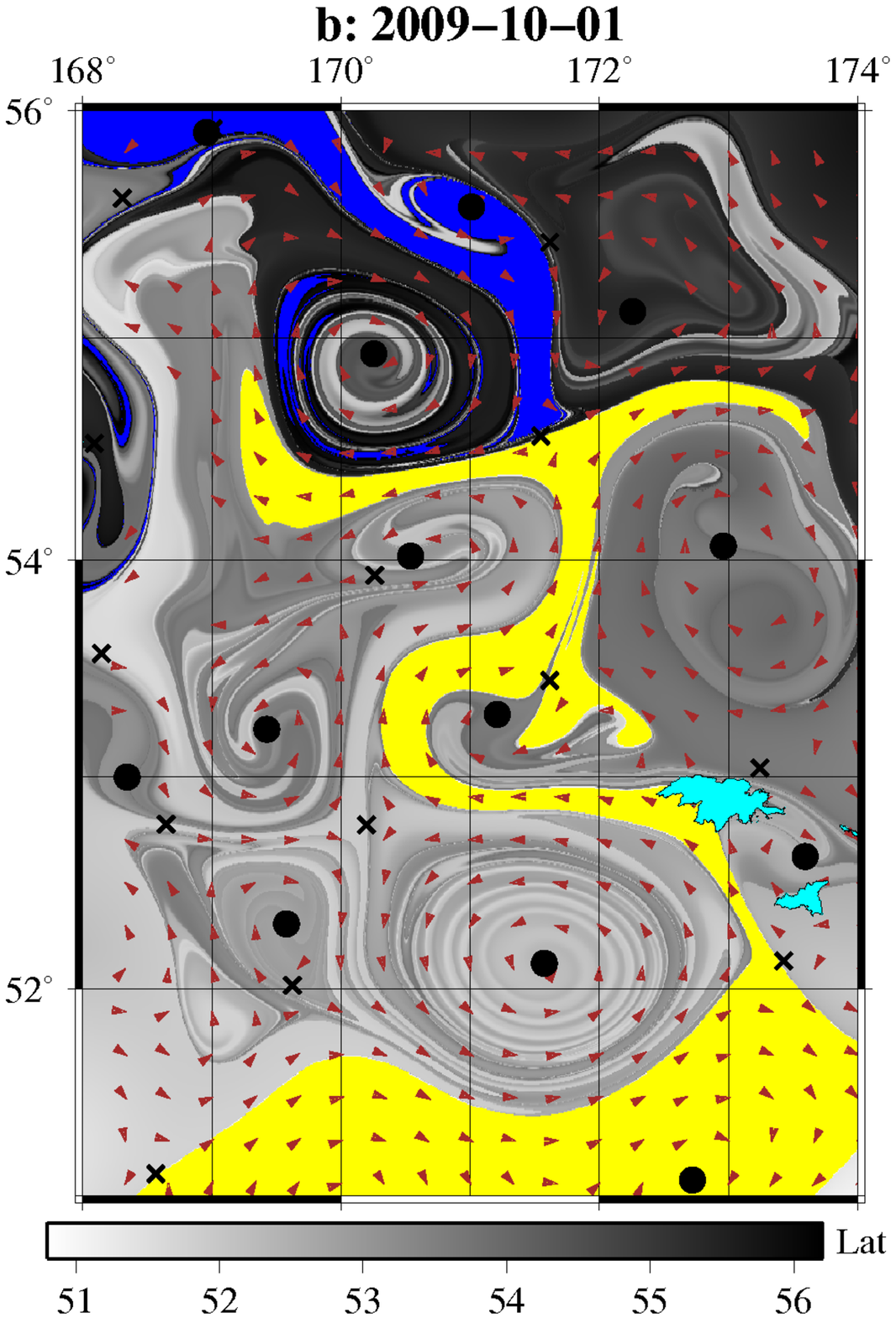}\\
\includegraphics[width=0.38\textwidth,clip]{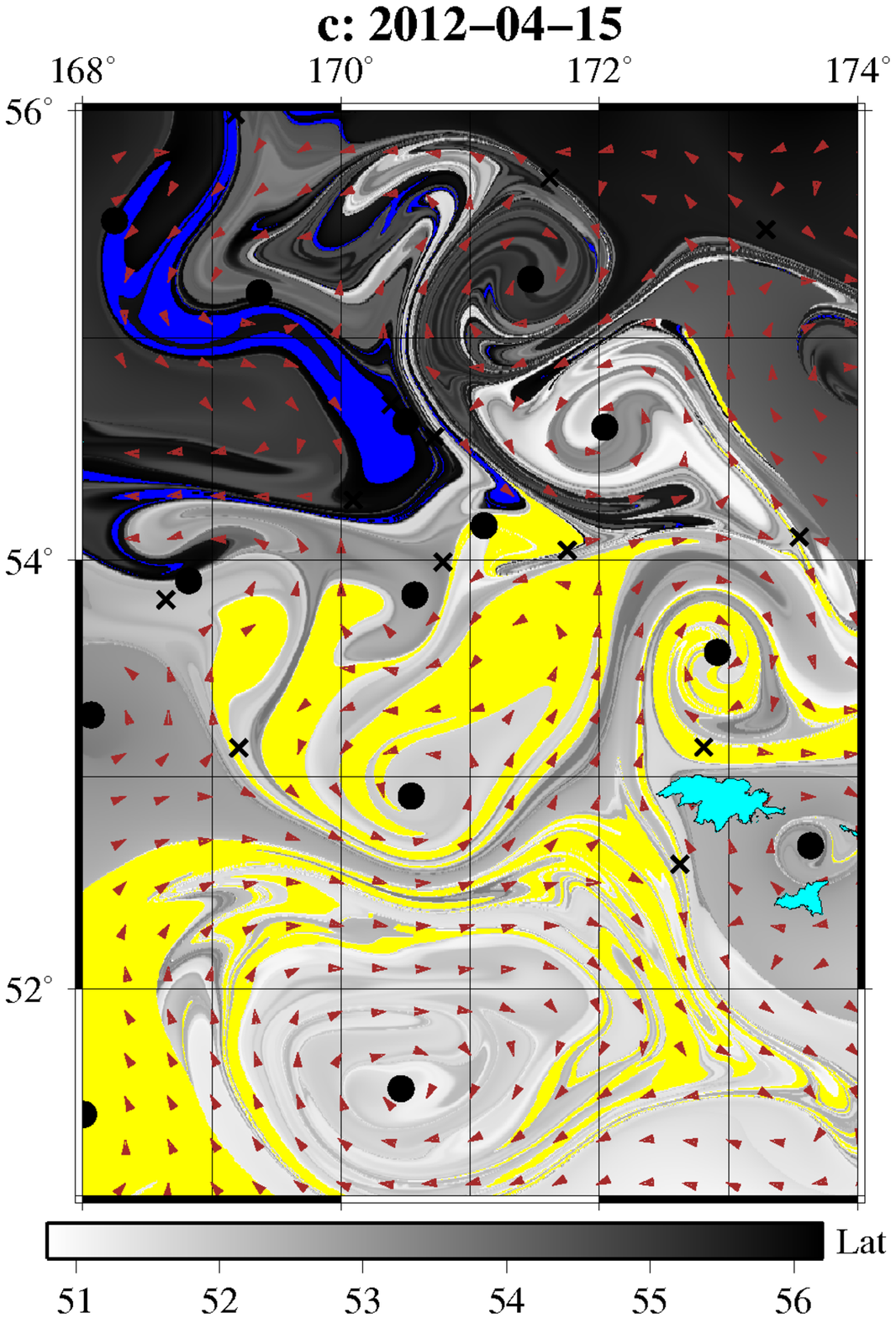}
\includegraphics[width=0.38\textwidth,clip]{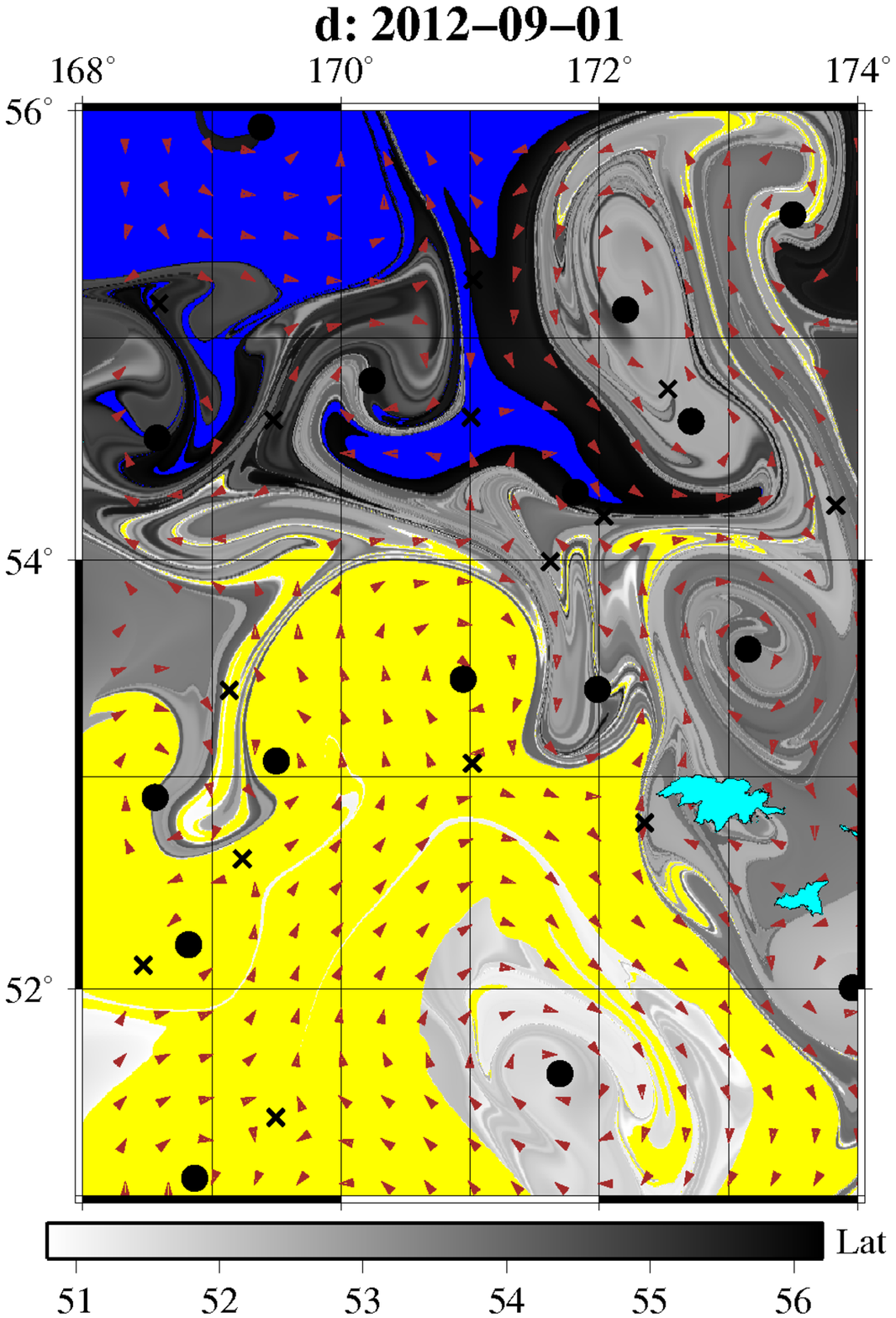}
\end{center}
\caption{Lagrangian latitudinal maps for four typical situations
demonstrating the impact of cyclonic and anticyclonic
eddies on the water transport across the NS. Mesoscale cyclonic eddy to the south of
the NS and a smaller scale cyclonic eddy to the north (a); triple eddy structure
consisting of the cyclonic mesoscale eddy, topographic anticyclonic eddy over the
Stalement Bank and a mesoscale anticyclonic eddy to the north of the NS (b);
the dipole vortex with the large anticyclonic eddy to the south of the NS and
a cyclonic one in the strait (c); the mesoscale anticyclone to the south of
the NS and the submesoscale topographic anticyclonic eddy over
the Stalement Bank (d). The color legend is the same as in Fig.~\ref{fig3}.}
\label{fig6}
\end{figure}

Water transport through the strait and its directions through different parts of the NS depend mainly
on the hydrological situation there. By analyzing Lagrangian maps in 1993--2013, we have found four typical
situations with the different number, scale and polarity of eddies to be present around and in the strait.
In Fig.~\ref{fig6} we illustrate those four cases on the latitudinal maps which demonstrate clearly not only
the eddies present in the area but the origin and history of water masses as well.

Figure \ref{fig6}a is an example of the situation with a mesoscale cyclonic eddy to the south of the NS
($x_0=170^\circ 30'$E, $y_0=52^{\circ}15'$N) and a smaller scale cyclonic eddy to the north of it
($x_0=170^\circ 30'$E, $y_0=53^{\circ}30'$N) with two elliptic points in their centers and
a hyperbolic point between them. Both the eddies are clearly seen in that panel dated by
August 1, 2009. It is seen that the southern cyclonic eddy enhances the flow from
the Bering Sea in the western part of the NR.
The ocean waters flow northward to the Bering Sea mainly through the central part of the NR (``light grey'' waters).
Two months later the mesoscale cyclone slightly shifted to the east
(Fig.~\ref{fig6}b on October, 1).
The submesoscale anticyclonic topographic eddy was formed over the Stalement Bank in the NS practically at the same
place as in August 2003 and a mesoscale anticyclonic eddy with the center
around $x_0=170^\circ 15'$E, $y_0=55^{\circ}$N
appeared to the north of it. This vortex structure cardinally changed the direction of transport
through the strait. The waters from the Pacific Ocean now
flow as a narrow white (yellow in the web version) meandering jet through the eastern part of the NS
along the unstable manifolds of that
vortex system encompassing all the eddies. In spring 2012 the situation was changed again.
The latitudinal map in Fig.~\ref{fig6}c on April 15, 2012 shows a dipole vortex with a large anticyclonic
eddy to the south of the NS ($x_0=170^\circ 30'$E, $y_0=51^{\circ}30'$N) and a cyclonic one in the strait
($x_0=170^\circ 30'$E, $y_0=53^{\circ}$N). That dipole organizes the flow in such a
way that the ocean waters (``white'' and ``yellow'' ones in
the printed and web versions, respectively) turn around both the eddies and between them blocking
the flux through the NS. Spiral-like sleeves of those waters reflect the process of trapping of
waters by the eddies from different latitudes for a period of integration (90 days).

The latitudinal map in Fig.~\ref{fig6}d has been computed on September 1, 2012 demonstrating the large scale
anticyclone to the south of the NS with the center around $x_0=171^\circ 30'$E, $y_0=51^\circ 30'$N and
the submesoscale topographic anticyclonic eddy over the Stalement Bank in the strait. In difference from
the situation in September 2003 (Fig.~\ref{fig3}c) there is no mesoscale anticyclone to the north of the
strait in September 2012. This twin vortex structure intensifies the flux to the Bering Sea through the
central part of the NS (``white'' and ``yellow'' waters in
the printed and web versions, respectively) without a significant outflow from the sea through its eastern part.

\begin{figure}[!htb]
\begin{center}
\includegraphics[width=0.9\textwidth,clip]{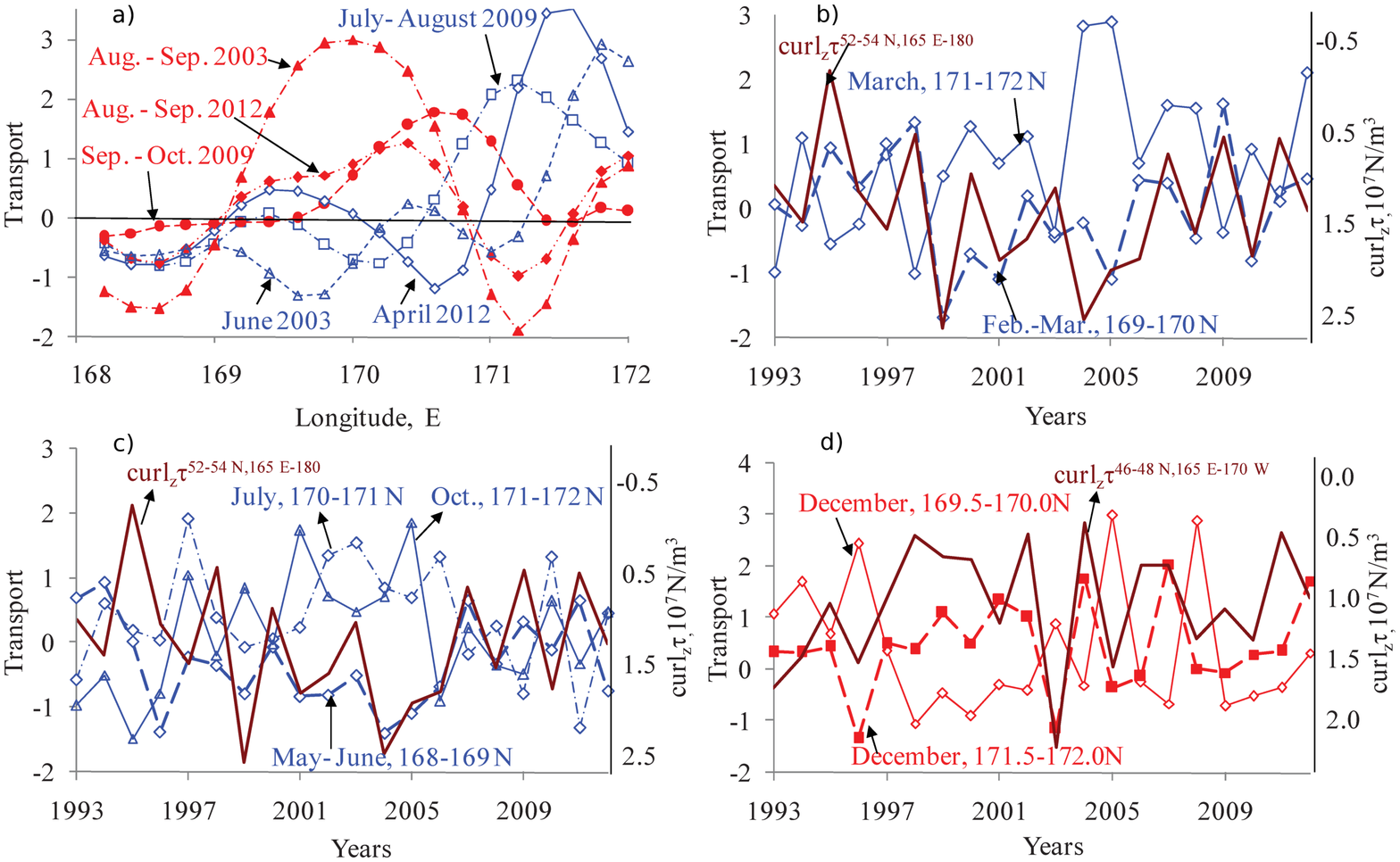}
\end{center}
\caption{Monthly averaged water transport in different parts of the
NS in 2003 (June, Aug.--Sep.), 2009 (July--Aug., Sep.--Oct.)
and 2012 (April, Aug.--Sep.) (a); year-to-year changes of the water transport
through the NS and the wind stress curl (Nov.--Mar.) in the NS area (b, c)
and central subarctic Pacific (d).}
\label{fig7}
\end{figure}
\begin{figure}[!htb]
\begin{center}
\includegraphics[width=0.6\textwidth,clip]{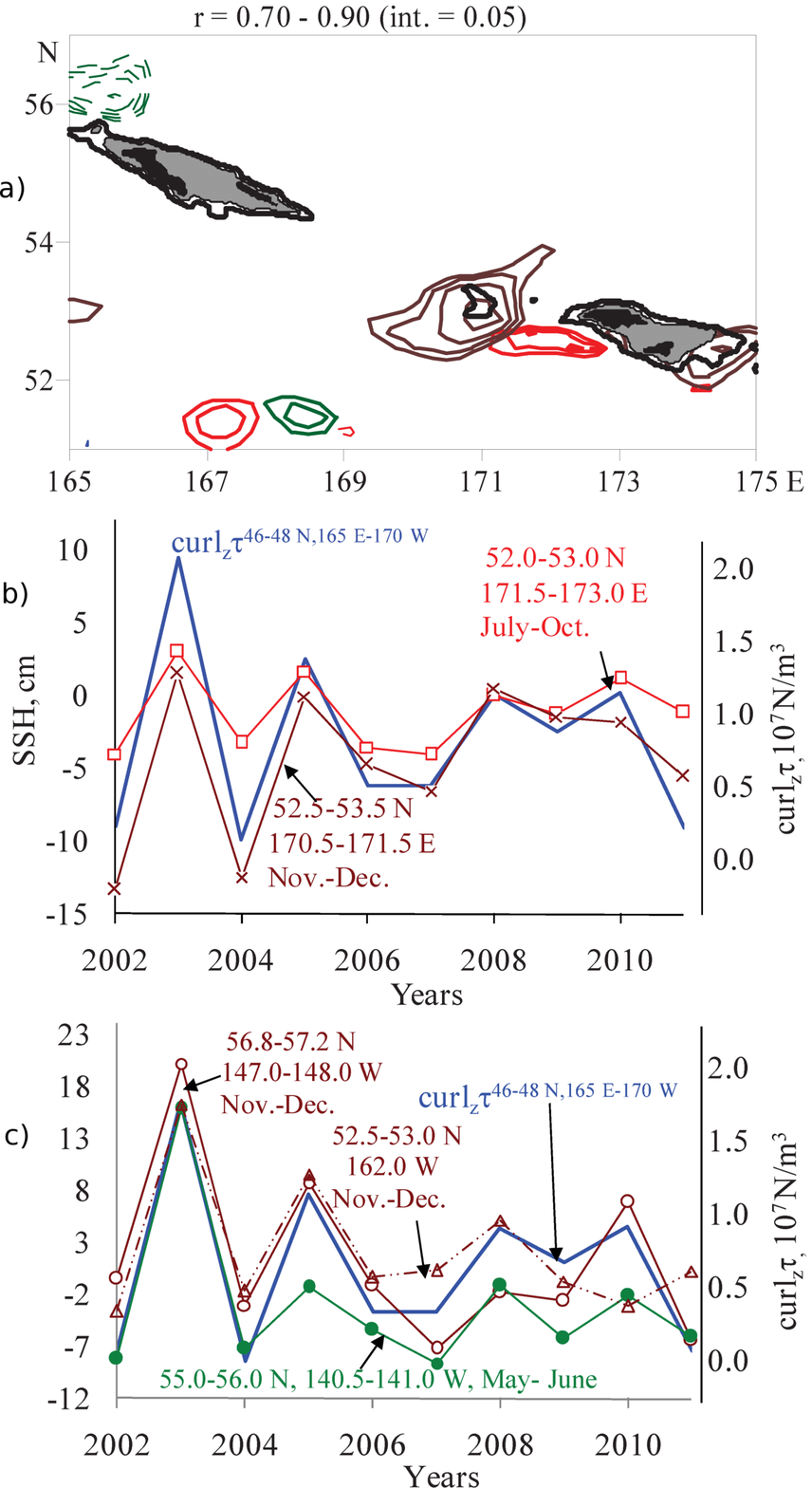}
\end{center}
\caption{Correlation coefficients between the temporal changes of the sea level
in the NS area and the wind stress curl (Nov.--Mar.) in the central subarctic Pacific (a);
year-to year changes of the sea level in the NS area (b) and at northern and western
boundaries of the subarctic Pacific (c) and the wind stress curl (Nov.--Mar.)
in the central subarctic Pacific (b--c).}
\label{fig8}
\end{figure}

By the eddy formation and interaction we can explain the results of the current mooring measurements
conducted in the NS from August 1991 until September 1992 \citep{Reed_Stabeno_1993}. Altimetry data,
provided by www.eddy.colorado.edu/ccar, demonstrate that flow pattern observed in the NS in June--August 1992
was similar to those in July--October 2003 (Fig.~\ref{fig3}). Interaction of the anticyclonic eddy located to
the south off NS with the Stalement Bank was led to the formation of the positive SSH
anomaly in the NS in June 1992. In late August 1992 this SSH anomaly moved northward into the Bering Sea
and was captured by the mesoscale anticyclonic eddy to the north of NS (Fig.~\ref{fig3}b). The formation of the positive
SSH anomaly in the Stalement Bank area could result in the increased northward flow in the central part and
southward  flow in the eastern part of the NS. This conclusion agrees well with the current mooring
observations (Reed and Stabeno \citep{Reed_Stabeno_1993} see their Figs.~\ref{fig7} and \ref{fig8}).
In June--August 1992 subsurface northward flow was intensified in the central part of the strait
(to the west of the Stalement Bank) and the flow reversed from the weak northward to the strong southward in
the eastern part of the NS (to the east of the Stalement Bank). In September 1992 (when the positive SSH
anomaly moved into the Bering Sea) the flow in the eastern side of the strait reversed from the southward
to the northward.

Until recently, the detailed study of the impact of cyclonic eddies on the volume flux between the Bering
Sea and the Pacific Ocean has not been conducted. Partly it's due to difficulties in cyclonic eddies
identification by using infrared and ocean color satellite images. The Lagrangian simulation helps to
shed light on that problem by computing daily latitudinal  maps in the area.
Lagrangian maps presented in Fig.~\ref{fig6} demonstrate  the effect
of mesoscale and submesoscale cyclonic and anticyclonic
eddies on the water transport through the NS in 2009 and 2012. During April--June 2012, the central and eastern parts of the
NS were under the influence of the mesoscale cyclonic eddy located to the south of the strait.
The formation of the submesoscale cyclonic eddy in the central part of the NS in April forced the
northward flow on the eastern side of the strait and the southward flow through the central part.

In October 2009 and September 2012, the strengthening of the mesoscale
anticyclonic eddies to the north of
and south of the NS led to the replacement of the submesoscale cyclone by
the submesoscale anticyclone
in the central part of the strait and thereby changed the flow pattern in the
strait. In October 2009,
the transport through central part of the strait changed its direction from
the southward to the northward.
During August 2012, the Alaskan Stream waters recirculated on periphery of the
anticyclone to the south of
the NS and did not penetrate into the strait.  In September 2012, due to the
appearance of a submesoscale
anticyclonic eddy in the central part of the NS, the water flux was directed
from the Pacific into the
Bering Sea through the central part of the strait, then the flow undergone a
retroflection and it supplied
into the Bering Sea on the eastern side of the NS (Fig.~\ref{fig6}).

The observations from the Argo buoys provide information about the water property in the NS. Two types of water
masses are observed in the strait area, relatively warm ($>4$~${}^{\circ}$C at 50--200~m) Alaskan Stream waters and
the western subarctic waters ($<4$~${}^{\circ}$C at 50--100~m). The Alaskan Stream is a continuation of the Alaska
Current originated in the NE Pacific. Western subarctic waters are supplied into the NS area on the eastern
periphery of the Western Subarctic (cyclonic) Gyre. In April 2012 the water
masses flowing  into the Bering Sea through the eastern part of the NS was
composed by the Alaskan Stream waters and western subarctic waters.
In September 2012, the main source
of the northward flow through the NS was the Alaskan Stream.

Fig.~\ref{fig7}a shows the flow reversals in the central and eastern parts of the NS in 2003, 2009 and 2012.
The reversals occur due to the relative positions of mesoscale eddies both to the north and to the south of
the NS and the cyclonic and anticyclonic submesoscale eddies in the strait (Figs.~\ref{fig3}, \ref{fig4} and \ref{fig6}).
In June 2003, July--August 2009 and April 2012 a submesoscale cyclone has pushed part of the Alaskan
Stream into the NS, causing a northward flow on the eastern side of the strait and a southward flow
in the central part of the strait. The formation of the submesoscale anticyclone during
August--September 2003, September--October 2009 and August--September 2012 modified the
flow through the NS, creating a strong northward flow through the central part of the
strait and a southward flow on its eastern side.

\subsection{Relationship between the wind stress curl and the eddy intensity}

The strength and position of the Aleutian Low pressure cell are main factors which determine the
oceanic dynamics in the northern North Pacific. The strong Aleutian Low and positive wind-stress
curl pattern over the northern North Pacific in November--March spin-up of the subarctic cyclonic
gyre (see e.g. \citep{Ishi_2005}). One may assume that that spin-up of the subarctic gyre results in
more eddy activity on its northern boundary (Alaskan Stream region). Our results demonstrate that
on the seasonal time scale the volume flux through the NS depends on wind stress curl
($\operatorname{curl}_{z} \tau=\partial{\tau_y}/\partial{x}-\partial{\tau_x}/\partial{y}$,
where $\tau_y$ and $\tau_x$ are respectively meridional and zonal
wind stress components) in the northern North Pacific (Figs.~\ref{fig7}b--d).
Increased wind stress curl over the NS in November--March enhances the southward transport on the
western side and a northward flow on the eastern side of the NS during winter and
spring ($r=-0.68$ and $0.61$, 1993--2012) (Figs.~\ref{fig7}b, \ref{fig7}c).
The cyclonic circulation is accompanied by decreased SSH in the NS area ($r=-(\text{0.75--0.90})$,
2002--2012).

In addition to local forcing, our results demonstrate that the wind stress curl in the central
part of the subarctic North Pacific ($46^{\circ}$N--$48^{\circ}$N, $165^{\circ}$E--$170^{\circ}$E)
determines the year-to-year changes in water transport through the NS during fall (Fig.~\ref{fig7}d).
Large positive wind stress curl in the subarctic Pacific corresponds to increased northward volume
flux in the central part of the NS ($r=0.66$) and vice versa for the eastern side of the strait
($r=-0.62$). The flow pattern in the NS during fall (October--December) is determined by the
strength of the anticyclonic eddy in the strait area. Analysis of the SSH time series at the
northern boundary of the subarctic gyre (the Alaskan Stream and the Alaska Current) revealed
that the year-to-year changes of the SSH in the anticyclonic eddies are related to the wind
stress curl ($r=0.85$--$0.89$, 2002--2012) (Figs.~\ref{fig8}a--c).
Spin up of the subarctic cyclonic gyre forced by the wind stress curl intensifies the
boundary currents, increases the sea level at the gyre
boundaries and probably enhances the anticyclonic eddy activity in the Alaskan Stream area.
Our conclusions are supported by the results of Ladd et al.~\citep{Ladd_Stabeno_OHern_2012}.
They indicated that the anticyclonic eddy activity along the eastern shelf-break of the
Bering Sea during the spring months is negatively correlated with the North Pacific Index,
a measure of the strength of the Aleutian Low in November--March. The proposed mechanisms
for the creation of anticyclonic eddies include instabilities and the interaction of the
boundary currents with topographic features.

\section{Conclusion}

The Lagrangian approach, based on the AVISO surface velocity field, is applied to study temporal
and spatial variations of the transport between the Bering Sea and the North Pacific. Our results
demonstrate that the eddy activity in the vicinity of the Near Strait has a strong impact on the
flow through the strait. Generation of anticyclonic and cyclonic submesoscale eddies in the Near
Strait leads to the flow reversals in central and eastern parts of the strait and westward
(or eastward) shift in the velocity core. In June 2003, July--August 2009 and April 2012 a
submesoscale cyclone has pushed part of the Alaskan Stream into the Near Strait, causing a
northward flow on the eastern side of the strait and a southward flow in its central part.
The formation of the submesoscale anticyclone during August--September 2003,
September--October 2009 and August--September 2012 modified the flow through the
Near Strait creating a strong northward flow through the central part and southward
flow on the eastern side.

On the interannual time scale, significant positive correlation ($r=0.72$) is diagnosed
between the Near Strait transport and the wind stress in winter. Increased southward component
of the wind stress decreases the northward water transport through the strait. Positive wind
stress curl over the strait area in winter--spring generates the cyclonic circulation and
thereby enhances the southward flow in the western part ($r=-0.68$) and northward flow in
the eastern part ($r=0.61$) of the Near Strait. During fall, the wind stress curl over the
northern North Pacific (November--March) are responsible for eddy activity in the Alaskan
Stream area and the flux pattern in the Near Strait.

\section*{Acknowledgments}

This work was supported  by the Russian Foundation for Basic Research
(project nos.~11--05--98542, 12--05--00452, and 13--05--00099).
The altimeter products were distributed by AVISO with support from CNES.

\bibliography{pap}
\end{document}